\documentstyle[aps]{revtex}

\begin{document}
\draft
\preprint{NYU-TH/99-01-03}  
\title{Reply to ``Comment on Correlation between\\ 
Compact Radio Quasars and Ultra-High Energy Cosmic Rays"\\
}
\author{Glennys R. Farrar$^a$ and Peter L. Biermann$^b$}
\address{$^a$Department of Physics, New York University, NY, NY 10003, USA\\
$^b$Max Planck Institut f{\"u}r Radioastronomie Auf dem H{\"u}gel 69, D-53121 Bonn, Germany\\
}

\date{\today}
\maketitle
\begin{abstract}

\end{abstract}
\pacs{
{\tt$\backslash$\string pacs\{\}} }


In ref. \cite{fb} we investigated the hypothesis that the highest
energy cosmic rays are created in and travel undeflected from an
extraordinary class of QSO's, capable on physical grounds of producing
the highest energy particles found anywhere in nature.  This 
{\it a priori} hypothesis was motivated by theories of cosmic ray
acceleration and the ansatz of a new, neutral, GZK-evading messenger
particle.  It is well known that many features of powerful AGN's are
not characteristics of every one of them and thus would not be
suitable markers (e.g., blazars only look like blazars when viewed
from a special direction).  The class of compact radio quasars (CQSOs)
is the only kind of quasar which holds any hope of at once
accelerating particles to very high energy, and at the same time 
converting them into another, possibly long-lived new particle by
interacting with material surrounding the AGN.  The distinctive radio
spectrum which provides an objective definition of the source class,
is in fact produced by interaction with the surrounding material and
is thus indicative of the conditions required.   
 
As pointed out by Elbert and Sommers\cite{elbert_sommers}, 3C147 is a
remarkable, uniquely suitable candidate source for the highest energy
cosmic ray event FE320, once one sets aside the GZK distance
limitation.  No matter what further restriction on characteristics of
the source class is made, 3C147 must be the source of FE320, given our
hypothesis that the source is an AGN.  Precisely for this reason, it
is correct to include it in the analysis independently of any
additional constraint ultimately imposed to remove background such as
the radio spectrum of a CQSO.   

Our energy cut was not arbitrary and was decided before examining the
events.  Clearly, one must provide some buffer against contamination
by mismeasured protons piled up just at the GZK limit.  Convoluting a
rapidly falling distribution with a gaussian measurement error means
the sample is preferentially drawn from the low-side of the
distribution.  With an $E^{-p}$ spectrum and $p = \{2,3,4\}$, an event
with nominal energy 1(2)-sigma above $E^{\rm cut}_{1(2)}$ has a true
energy lower than $E^{\rm cut}_{1(2)}$ \{37\%, 40\%, 42\%\} (\{5.0\%,
5.2\%, 5.5\%\}) of the time.  In the analysis reported in \cite{fb} we
required that the event should have an energy of at least $8 ~10^{19}$
eV plus 1-sigma.  This was motivated by the result of ref. \cite{cfk}
that protons from nearby AGN's (~100 Mpc) have energies degraded to the $ 
5-8 ~10^{19}$ eV range by interaction with CMBR, independently of
their initial energy.  In analyses of future data, we plan to impose
an additional cut, that the nominal energy of the event be at least $5
~10^{19}$ eV plus 2-sigma.  For distant sources this would produce a
$\sim 95\%$ clean sample.  The existing set of events passes this cut
as well. These cuts are soft in the sense that the $8 ~10^{19}$ eV
starting point for defining $E^{\rm cut}_1$ is not precise -- $ 7~ {\rm
or}~ 9 ~10^{19}$ eV could also have been chosen and would have
provided weaker or stronger background rejection.

Ag110 has a central energy value of $1.1 \pm 0.3 ~10^{20}$ eV and thus 
satisfies both cuts.  Hoffmann adds a third significant figure to the
energy and claims the event should be excluded from the analysis
since 1.10 - 0.33 = 0.77.  We disagree since the cut energy was
defined at 1 significant figure.  However our result is robust:
excluding Ag110, the CQSO $\chi^2$ probability is unchanged (0.53),
the testQSO probability remains small ($4 ~10^{-7}$), and the
probability that randomly distributed QSO's would produce as low or
lower $\chi^2$ is still small (0.016).  The possible contention over
whether or not to include the event underlines the point emphasized in
our paper: a definitive resolution of the issue requires more and more
precise data.    

Table I contains three typos: the RA error for HP120 should be 2.7
deg;  $\Delta \Omega$ for FE320 and HP120 should be 1.9 and 6.7
deg$^2$ respectively.  Correct values of all parameters were used in
the analysis.  Note that $\Delta \Omega$ is only a figure of merit and
does not enter the analysis.  The conventional difference between how
conical errors and rectilinear errors are quoted is correctly
incorporated in the formulae (see Eq. 2 and text below) but not
explicitly discussed due to space limitations. 

Hoffman's Comment prompted us to analyze the three events which come
closest to passing our cuts\cite{watson:HEPiN}, keeping in mind that
with the energies and errors of these three CR events and the falling
cosmic ray spectrum, it is likely that at least one of them should be a
proton because the probability that one of the three has a true energy
below $5~ 10^{19}$ eV is \{40\%,60\%\}, for $\{p=2,~4\}$, and at the
source protons far outnumber the messengers for a given energy.
Rather than diminishing the evidence in favor of the ansatz, the
characteristics of these three events fit the CQSO hypothesis very
well.  Ya110 (RA = $75.2 \pm 10$ deg, Dec = $45.5 \pm 4$
deg\cite{halzen}) would have the same source that produced the highest
energy event FE320 (3C147), with a $\chi^2$ residual of 2.0 for 2
degrees of freedom; the CQSO (0.53) and random background probability
(0.0058) are essentially unchanged by including Ya110.  There is an
excellent CQSO candidate for HP105 (B3 1325+436); it is an archtypal
CQSO and has a $\chi^2$ residual of 0.46.  Adding this event in the
analysis {\it decreases} the random background probability to 0.003
(0.0029 with Ya110 and 0.0028 without).  There is no good candidate
for HP102 within a cone of radius 5 deg so it is interpreted as a
deflected proton.   

We reiterate that whether the random background probability is 3\% or
0.3\% is not the essential point -- either value is low enough that
in coin-tossing experiments one would be surprised by it, but not low
enough for a statistical fluctuation to be excluded.  Our main message
is that future detectors should aim for the best possible position
resolution in order to settle this important question.



\begin{thebibliography}{10}

\bibitem{cyhoffman}
C.~Hoffman.
\newblock astro-ph/9901026.

\bibitem{fb}
G.~R.~Farrar and P.~L.~Biermann.
\newblock {\em Phys. Rev. Lett.}, 81:3579, 1998.

\bibitem{cfk}
D.~J.~Chung, G.~R. Farrar, and E.~W. Kolb.
\newblock {\em Phys. Rev.}, D57:4606, 1998.

\bibitem{watson:HEPiN}
A.~Watson.
\newblock Proceedings Snowmass workshop, p 126, 1996.

\bibitem{halzen}
F.~Halzen et~al.
\newblock {\em Astroparticle Phys.}, 3:151, 1995.

\bibitem{elbert_sommers}
J.~Elbert and P.~Sommers.
\newblock {\em Astrophys. J.}, 441:151--161, 1995.

\end{thebibliography}

\end{document}